\documentclass[aps,prl,amsfonts,amsmath,nofootinbib,preprint,tightenlines,longbibliography]{revtex4-1} 
\pdfoutput=1

\usepackage{graphicx}

\def\be{\begin{equation}}
\def\ee{\end{equation}}
\def\bea{\begin{eqnarray}}
\def\eea{\end{eqnarray}}
\def\bml{\begin{subequations}}

\def\elea{\end{eqnarray}\end{subequations}}
\def\ogw{\Omega_{\text{gw}}}

\begin{document}

\title{New limits on cosmic strings from gravitational wave observation}

\author{Jose J. Blanco-Pillado}
\email{josejuan.blanco@ehu.es}
\affiliation{Department of Theoretical Physics, University of the Basque Country, Bilbao, Spain}
\affiliation{IKERBASQUE, Basque Foundation for Science, 48011, Bilbao, Spain}

\author{Ken D. Olum}
\email{kdo@cosmos.phy.tufts.edu}
\affiliation{Institute of Cosmology, Department of Physics and Astronomy,
Tufts University, Medford, MA 02155, USA}

\author{Xavier Siemens}
\email{siemens@gravity.phys.uwm.edu}
\affiliation{Center for Gravitation, Cosmology, and Astrophysics,\\
  Department of Physics, University of 
  Wisconsin--Milwaukee, Milwaukee, WI 53201, USA}

% PhySH concepts: Gravitational wave sources, Cosmic strings & domain walls
% Gravitational wave detection

\begin{abstract}

We combine new analysis of the stochastic gravitational wave
background to be expected from cosmic strings with the latest pulsar
timing array (PTA) limits to give an upper bound on the energy scale
of the possible cosmic string network, $G\mu < 1.5\times 10^{-11}$ at
the 95\% confidence level.  We also show bounds from LIGO and to be
expected from LISA and BBO.

Current estimates for the gravitational wave background from
supermassive black hole binaries are at the level where a PTA
detection is expected.  But if PTAs do observe a background soon, it
will be difficult in the short term to distinguish black holes from
cosmic strings as the source, because the spectral indices from the
two sources happen to be quite similar.

If PTAs do not observe a background, then the limits on $G\mu$ will
improve somewhat, but a string network with $G\mu$ substantially below
$10^{-11}$ will produce gravitational waves primarily at frequencies
too high for PTA observation, so significant further progress will
depend on intermediate-frequency observatories such as LISA, DECIGO
and BBO.

\end{abstract}

\maketitle

Our universe may contain a network of cosmic strings arising as
topological defects in unified field theories or as fundamental strings
(or 1-dimensional D-branes) in string theory \cite{Vilenkin:2000jqa,Dvali:2003zj,Copeland:2003bj}.  If so, the best hope
for detecting this network is the observation of gravitational waves
from oscillating string loops.  Correspondingly, the strongest limits
on such a network arise from non-observation of such gravitational
waves.

Gravitation wave observations may detect a cosmic string network either
through bursts of radiation emitted at cusps or through the stochastic
background composed of radiation from all loops existing through
cosmic history.  Bursts were discussed in
Refs.~\cite{Damour:2000wa,Damour:2001bk,Damour:2004kw,Siemens:2006vk};
here we will concentrate on the stochastic background.

Current work \cite{Blanco-Pillado:2017oxo} has advanced our understanding of
the expected gravitational wave background spectrum, taking into
account all known effects, except that Ref.~\cite{Blanco-Pillado:2017oxo}
modeled gravitational back reaction by convolution instead of exact
calculation of the back-reaction effect on each loop.  It is thus
appropriate to update existing observational bounds
\cite{Siemens:2006yp,DePies:2007bm,Olmez:2010bi,Sanidas:2012ee,Binetruy:2012ze,Kuroyanagi:2012wm,Blanco-Pillado:2013qja}
on cosmic string network properties.  

Cosmic strings are classified by their tension or energy per unit
length $\mu$.  The gravitational effects of strings depend on the
product $G\mu$, where $G$ is Newton's constant.  We will work in units
where $c = 1$, so $G\mu$ is a dimensionless number.
Early models of strings considered $G\mu\sim 10^{-6}$, which is what
one might expect for symmetry breaking at the grand unification scale.
At such values, strings could be the cause of structure formation, but
they were ruled out long ago as a primary source of large scale structure
perturbations in the universe by cosmic microwave background (CMB)
observations.  Current limits from CMB observations give $G\mu\lesssim 10^{-7}$
\cite{Ade:2013xla}, but pulsar timing observations give a much stronger
limit, as we will discuss.

We consider here the usual model of local strings, with no couplings
to any light particle except the graviton.  At any given time the
cosmic string ``network'' consists of infinite strings and a
distribution of loops of various sizes.  Loops are formed by
reconnection of long strings with themselves.  They oscillate
relativistically and eventually lose their energy to gravitational
radiation.

In a uniform cosmological epoch, e.g, the radiation era between the
time of electron-positron recombination and matter-radiation equality,
long strings form a scaling network, in which the energy density of
strings maintains a fixed ratio to the critical density by redshifting
as matter in the matter era or radiation in the radiation era.  This
is possible because the excess energy which would otherwise
accumulate is transferred into the production of loops.

The loop population is diluted like matter, even in the radiation era.
As a result, radiation-era loops are greatly enhanced over long
strings.  With current bounds on $G\mu$, relic loops from the
radiation era always dominate over loops produced more recently.  The
effect on the gravitational wave background of loops formed in the
matter era is negligible.

To compute the stochastic background, we integrate the emission of
gravitational waves at each redshift $z$, transferred to the present
accounting for redshifting of the waves and dilution of the energy in
the subsequent cosmological evolution.  In a uniform radiation era,
there is a coincidence which leads to a flat spectrum in $\ogw$ with
amplitude that depends on the number density of loops and the total
gravitational power of a loop, $\Gamma G \mu^2$, but not on the shape 
of the emission spectrum.

The high frequency background today consists almost entirely of
gravitational waves emitted in the radiation era and thus participates
in this flat spectrum.  However, there is an important correction
because at early times the number of relativistic degrees of freedom
was changing, mainly at the times of the quark-hadron transition and
electron-positron annihilation.  These changes in the cosmological
kinematics lead to a decline in the spectrum towards higher
frequencies.

At lower frequencies, the spectrum is increasingly dominated by
gravitational waves emitted in the matter era (from relic
radiation-era loops).  Again the different kinematics lead to an
enhancement in the background.  At still lower frequencies, the
background spectrum falls rapidly, because there are few loops large
enough to emit such long waves.

Here we use the results of Ref.~\cite{Blanco-Pillado:2017oxo}, which
extracted loops from simulations, smoothed them to model the change of
shape due to gravitational back reaction, computed the gravitational
power from each loop and propagated the resulting gravitational waves
to find the present-day stochastic background.  Many of these details
make only a small difference to the gravitational spectrum, but there
are a few points that must be handled correctly.  Most importantly, it
is not correct to compute the loop density by applying energy
conservation to the long string network, because the great majority of
the energy appears as the kinetic energy of small, rapidly moving
loops, which contribute little to the gravitational spectrum
\cite{Blanco-Pillado:2013qja}.  For more information, see
Ref.~\cite{Blanco-Pillado:2017oxo}.

\begin{figure}
\begin{center}
\includegraphics[width=6.3in]{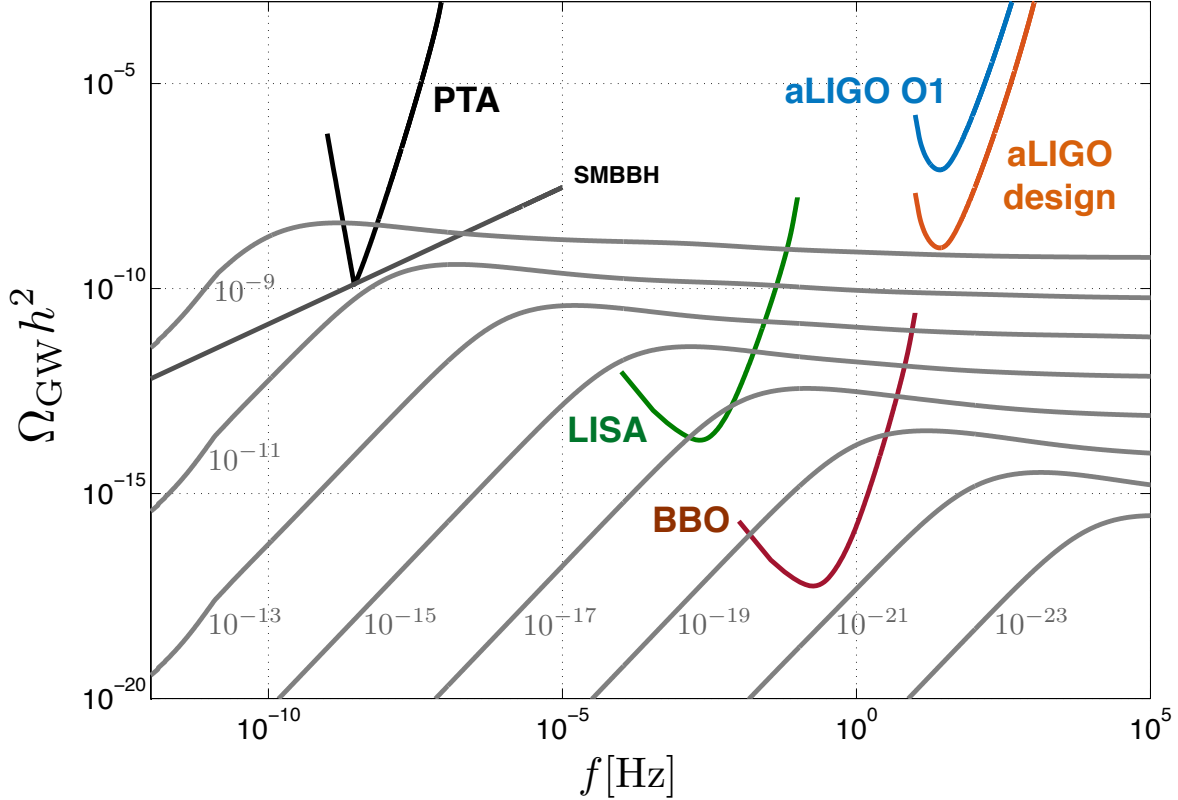}
\caption{Stochastic gravitational wave backgrounds compared with
  present and future experiments.  The grey lines show the background
  from cosmic strings with the indicated energy scales $G\mu$.  The
  straight black line is the largest allowable background from
  SMBBH. The remaining curves show the sensitivities of the various
  instruments.}
\label{fig:gwOmega}
\end{center}
\end{figure}

\begin{figure}
\begin{center}
\includegraphics[width=6.3in]{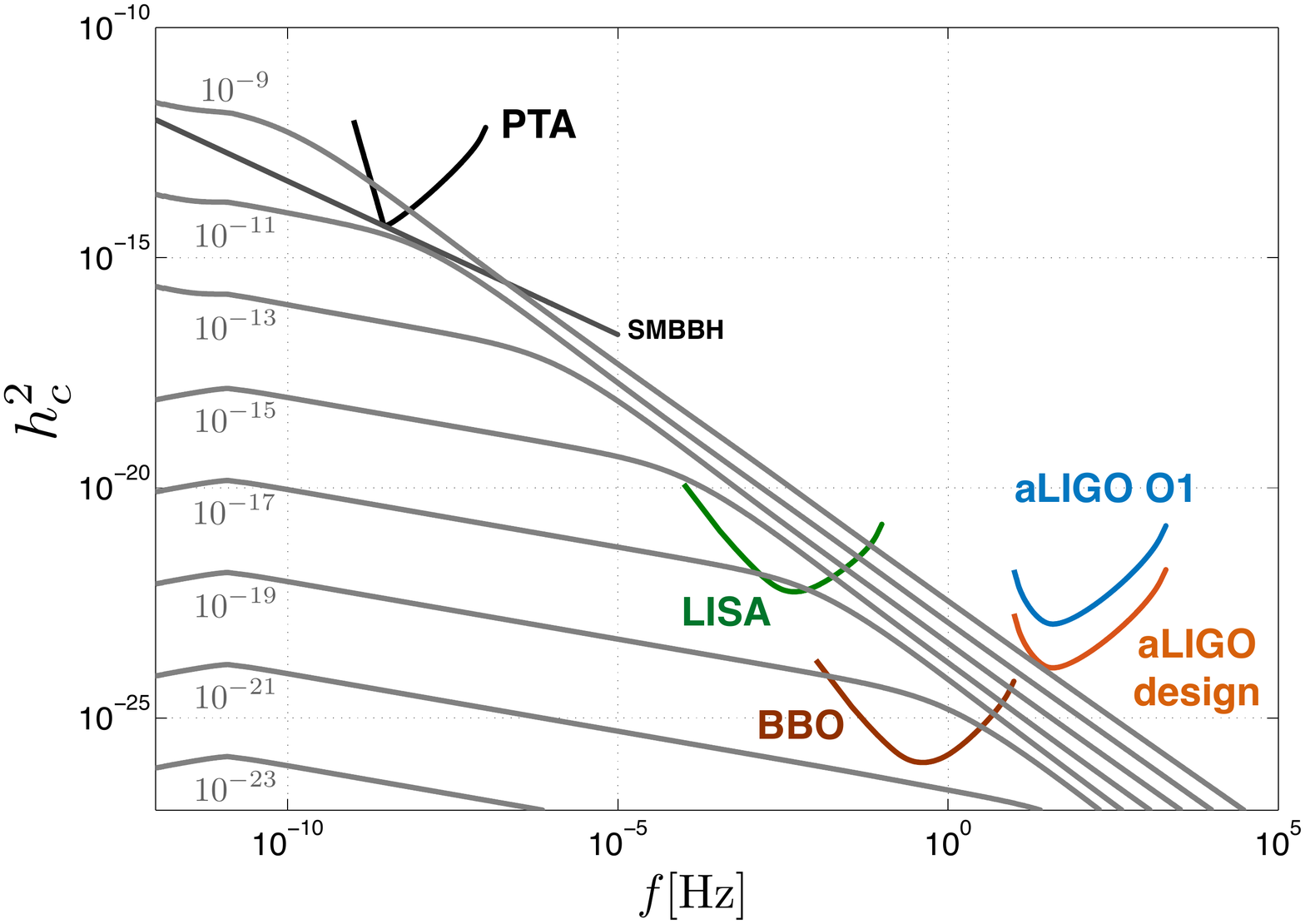}
\caption{As in Fig.~\ref{fig:gwOmega} but for characteristic strain
  $h_c$ (not including the solid angle and polarization sensitivity
  factor ($\sqrt{5}$ for LIGO or VIRGO) conventionally used by
  interferometers).}
\label{fig:gwhc}
\end{center}
\end{figure}

Figures \ref{fig:gwOmega} and \ref{fig:gwhc}
show the cosmic string gravitational-wave spectra for cosmic string
tensions in the range of $G\mu=10^{-23}$ to $10^{-9}$, along with a
sample of current and future gravitational wave experiments. We show
the current upper limit spectra for the LIGO O1
run~\cite{TheLIGOScientific:2016dpb}, advanced LIGO (aLIGO) at design
sensitivity assuming 1 year of integration, and the best PTA limit
(from the Parkes PTA \cite{Shannon:2015ect}) along with the spectrum
produced by supermassive binary black holes (SMBBH) with characteristic
amplitude $10^{-15}$, which is the largest currently allowed by that
limit \cite{Shannon:2015ect}.  We include curves for
LISA~\cite{Audley:2017drz} and BBO~\cite{Corbin:2005ny,Crowder:2005nr}
assuming 5 years of observation.

The instrumental curves are in the ``power-law integrated'' form
\cite{Thrane:2013oya}, and were calculated using the publicly
available codes used in Ref.~\cite{Thrane:2013oya}.  In the case of
LIGO and PTAs, this means that any power-law spectrum tangent to the
instrumental curve has been ruled out at the 95\% confidence level.
For future experiments, the instrument will be able to exclude such
spectra at 95\% confidence level 95\% of the time, if there is no
signal.  Either a Bayesian or a frequentist analysis can be used to
derive these curves, with little difference in the result.

With current limits on $G\mu$, the peak of the background is close to
the nHz frequencies to which pulsar timing arrays are most sensitive.
However, if $G\mu$ is significantly below these limits, PTA
frequencies will lie on the falling edge of the spectrum at low
frequencies, and discovery or further constraint will become more
difficult.  At that point, other instruments such as LISA,
\mbox{(B-)DECIGO} \cite{Sato:2017dkf}, and BBO will have much greater
ability to study cosmic string gravitational wave backgrounds.

LIGO frequencies are to the right of the peak and suffer from the
decrease due to changes in the number of degrees of freedom at early
times, and the sensitivity of LIGO to measure $\Omega$ is
significantly worse than PTAs.  Thus LIGO is not expected to be
competitive for this purpose.  The proposed Einstein Telescope
\cite{Punturo:2010zz} would operate at frequencies generally similar
to LIGO with about 100 times the sensitivity in $\ogw$.  From
Fig.~\ref{fig:gwOmega}, we see that this gives only slightly more
reach for cosmic string backgrounds that current PTAs, and is not
competitive with LISA.

\begin{figure}
\begin{center}
\includegraphics[width=6.3in]{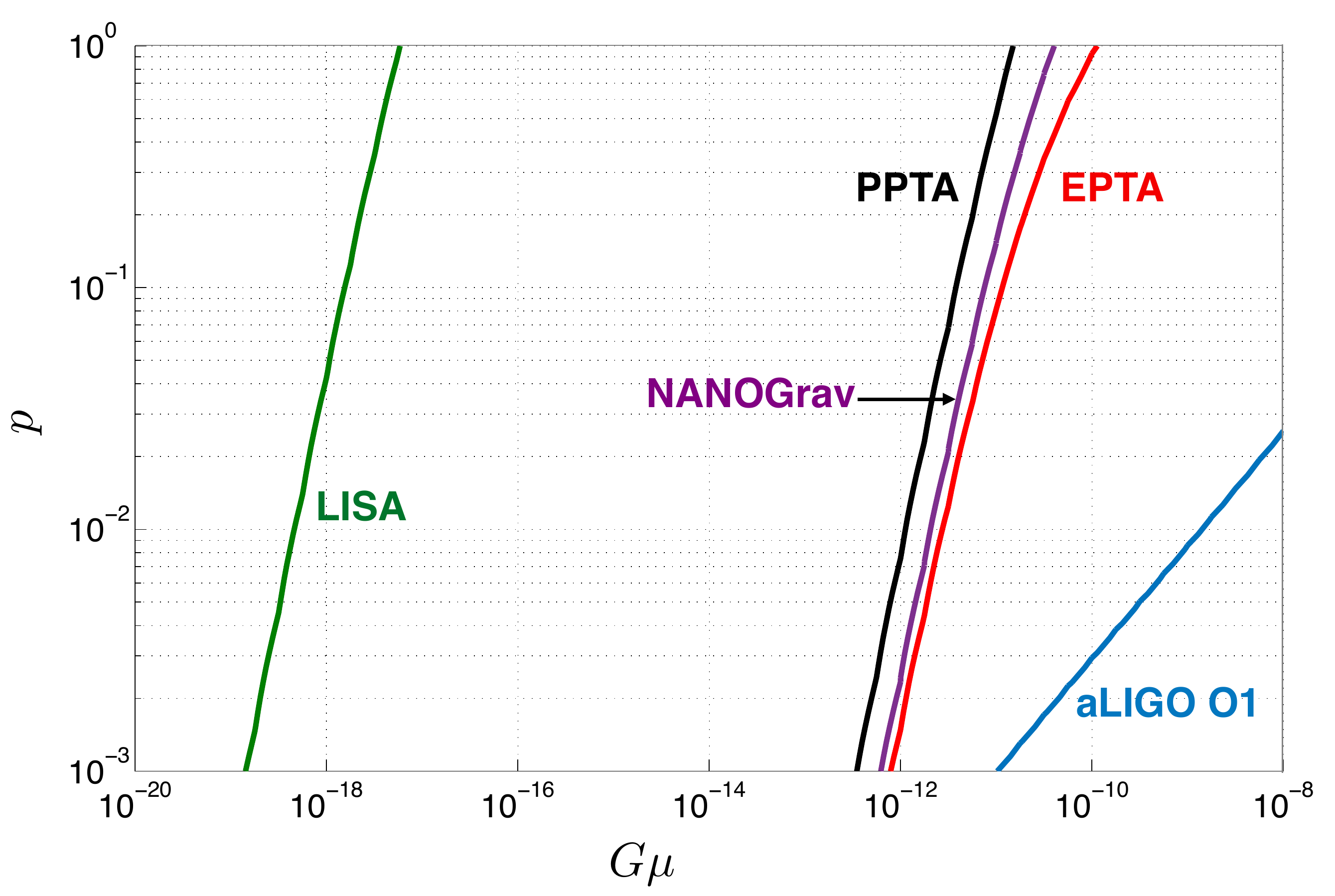}
\caption{Excluded regions in the $G\mu$-$p$ plane from present and
  potential future observations, by the instruments labeled.  The
  excluded area is to the right of each curve.}
\label{fig:exclusion}
\end{center}
\end{figure}

Figure \ref{fig:exclusion}
shows current 95\% confidence limits on $G\mu$ from present and future
experiments (in the absence of a detection).  These constraints are
derived as follows.  LIGO~\cite{TheLIGOScientific:2016dpb} gives a
constraint on the energy density of a flat spectrum of gravitational
waves, $\Omega < 1.7 \times 10^{-7}$ in the frequency band 20--86 Hz.
Since the cosmic string spectrum is close to flat in this band, this
constraint can be used directly.  Pulsar timing experiments report
constraints on $\ogw$ at a specific frequency, the one in which the
observations are most sensitive.  This constraint applies not only to
a flat spectrum but also to a wide range of power laws, and the
effects of the period of observation are taken into account as in
Ref.~\cite{Thrane:2013oya}.  The 95\% confidence limits are $\Omega
h^2 < 1.2 \times 10^{-9}$ at frequency $5\times 10^{-9}$~Hz for the
EPTA~\cite{Lentati:2015qwp}, $\Omega h^2 < 4.2 \times 10^{-10}$ at
frequency $3.3\times 10^{-9}$~Hz for
NANOGrav~\cite{Arzoumanian:2015liz}, and $\Omega h^2 < 10^{-10}$ at
frequency $2.8\times 10^{-9}$~Hz for the PPTA~\cite{Shannon:2015ect,
  Lasky:2015lej}.  We then simply find the $G\mu$ at which each
constraint is saturated.  For LISA we find the $G\mu$ that would lead
to a 95\% chance of detection in 5 years of observation, using the
techniques and publicly available codes used in~\cite{Thrane:2013oya},
applied to the predicted cosmic string background spectra.

In Fig.~\ref{fig:exclusion} we have a plotted the limits on $G\mu$
against possible intercommutation probability $p$, using the
conventional assumption that $p < 1$ simply increases the network
density and thus the gravitational wave background by factor $1/p$.
However, we note that while long string reconnection is the same in a
denser network with lower $p$, loop production depends on strings
reconnecting with themselves, which is not affected by the overall
density.  Thus we feel that the nature of low-$p$ networks may not be
well understood.

Among present experiments, the strongest limit comes from the Parkes
PTA \cite{Shannon:2015ect,Lasky:2015lej}, which gives
$G\mu < 1.5\times 10^{-11}$ for $p=1$.  NANOGrav results give 
of $G\mu < 4.0\times 10^{-11}$ \cite{Arzoumanian:2015liz}, while EPTA
\cite{Lentati:2015qwp} gives $G\mu < 1.1\times 10^{-10}$.  LISA could in
principle set a limit of $5.8\times 10^{-18}$, but here we have not
considered effect of foregrounds and other backgrounds.

These limits apply to string networks that lose their energy primarily
by gravitational waves.  Global strings would evade it, as would
strings that couple directly to some moduli field or the Higgs
\cite{Sabancilar:2009sq,Hyde:2013fia,Long:2014lxa} or have other
mechanisms for emitting energy into massless
particles\footnote{Refs.~\cite{Hindmarsh:2008dw,
    Bevis:2010gj,Hindmarsh:2017qff} simulated cosmic string networks
  in lattice field theory for the Abelian Higgs model and found energy
  going almost entirely into the high-mass particles of the string
  fields. However, it is very hard to see how such results could be
  applicable to a realistic situation where there is a vast difference
  between the cosmological and string scales, which suppresses
  any massive radiation from the strings (See for example
  Refs.~\cite{Olum:1999sg,Moore:1998gp}).}.
\pagebreak

Strings were first proposed as an explanation for structure formation,
with symmetry breaking at the grand unification scale giving $G\mu
\sim 10^{-6}$ \cite{Zeldovich:1980gh,Vilenkin:1981iu}.  Such models
were ruled out in the late 1990s by the acoustic peaks in the CMB
power spectrum \cite{Albrecht:1997nt,Lange:2000iq,Albrecht:2000hu}.
In the two decades since then, gravitational wave background limits
have lowered the maximum possible $G\mu$ by five orders of magnitude
to $1.5\times 10^{-11}$.  At this level and below, we will not see any
effects in the CMB, nor will we discover strings by gravitational
lensing (except perhaps microlensing in certain models
\cite{Chernoff:2014cba}). Other possible gravitational effects of
strings \cite{Khatri:2008zw,Shlaer:2012rj,Barton:2015zra} will also
be significantly limited by this bound.

Usual models of the evolution of black holes from merging galaxies
predict that an SMBBH background should already have been observed
\cite{Shannon:2015ect}.  Thus it seems likely that such a background,
if nothing else, will soon be detected.  If and when a gravitational
wave background is observed, it will be important to distinguish
cosmic strings from SMBBH as the source.  Since both give Gaussian
backgrounds,\footnote{We may be able to distinguish a cosmic string
  background from one produced by SMBBHs via anisotropy
  measurements~\cite{Mingarelli:2013dsa, Cornish:2015ikx}; further
  work is needed to establish whether this is a viable possibility.}
the only distinguishing feature is the spectrum.  For black holes,
$\Omega\sim f^{2/3}$ up to a cutoff.  For cosmic strings, the spectrum
is more complicated, with a rising power $\Omega\sim f^{3/2}$, then a
peak, followed by a small decline and then a plateau.  At the current
limit, $G\mu \sim 1.5 \times 10^{-11}$, PTA frequencies lie just to
the left of the peak, and unfortunately the spectral index there is
quite close to the $f^{2/3}$ of black hole binaries, as shown in
Fig.~\ref{fig:gwOmega}. Thus if a signal is detected at about this
level, it will be difficult to attribute it definitively to black
holes or strings.  On the other hand, even a small increase in PTA
sensitivity without detection will break this degeneracy by further
limiting the maximum $G\mu$, and so pushing the peak to higher
frequencies.  Then PTA frequencies will lie in the $f^{3/2}$ region of
the cosmic string spectrum.  In any event, LISA or (B-)DECIGO will
make this distinction very clearly.

\vspace{10pt}

We thank Joe Romano and Neil Cornish for useful conversations and Joe
Romano and Eric Thrane for making public the codes used in
Ref.~\cite{Thrane:2013oya}.  X. S. was supported in part by the
National Science Foundation under Grant No.\ 1430284 (NANOGrav Physics
Frontiers Center) and No. 1607585.  K. D. O. was supported in part by
the National Science Foundation under Grant No.\ 1213888, No.\ 1213930,
No.\ 1518742, and No.\ 1520792.  J. J. B.-P.  was supported in part by
the Basque Foundation for Science (IKERBASQUE), the Spanish Ministry
MINECO grant (FPA2015-64041-C2-1P) and Basque Government grant
(IT-979-16).

The data from Ref.~\cite{Blanco-Pillado:2017oxo}, used to make the
cosmic string curves in Fig.~\ref{fig:gwOmega}, are available at
\url{http://cosmos.phy.tufts.edu/cosmic-string-spectra/}.

\bibliography{comparison}

\end{document}